\documentclass[twocolumn,dvipsnames,11pt]{scrartcl}
\usepackage[colorlinks,pdfusetitle,urlcolor=blue,citecolor=blue,linkcolor=blue,bookmarksnumbered,plainpages=false]{hyperref}
\usepackage{amsmath}
\usepackage{amssymb}

\usepackage{eurosym}
\usepackage{booktabs}
\usepackage{color}
\usepackage{hhline}
\usepackage[top=1.5in,bottom=1.5in,left=0.7in,right=0.7in]{geometry}
\usepackage{tcolorbox}
\usepackage{bm}
\usepackage{nicefrac}
\usepackage{wrapfig}
\usepackage{gensymb}
\usepackage[bf]{caption}
\newcommand{\vect}[1]{\bm{#1}}
\newcommand{\ten}[1]{\mbox{\textbf{%\textit
{\textsf{#1}}}}}

\newcommand{\trace}{\operatorname{tr}}

\newcommand{\dif}{\mathrm{d}}
\newcommand{\mi}{\mathrm{i}}
\newcommand{\me}{\mathrm{e}}

\newcommand{\cavityDecay}{\kappa}
\newcommand{\totalLoss}{\Gamma_{\downarrow}}
\newcommand{\totalGain}{\Gamma_{\uparrow}}
\newcommand{\pumpRate}{\gamma_{\uparrow}}
\newcommand{\lossRate}{\gamma_\downarrow}
\newcommand{\stimEmission}{\gamma_{\downarrow\nu}}
\newcommand{\stimAbsorp}{\gamma_{\uparrow\nu}}

\newcommand{\peakSplitting}{\Delta \Omega}
\newcommand{\dyeCenter}{\Omega_0}
\newcommand{\dyePeakBroadening}{\delta}
\newcommand{\NMol}{M}
\newcommand{\photonN}{N}

\usepackage{bm}
\usepackage{braket}

\usepackage{wrapfig}

\usepackage[style=phys,url=false,isbn=false,doi=false,maxbibnames=3]{biblatex}
\AtEveryBibitem{\clearfield{month}}
\AtEveryBibitem{\clearfield{title}}
\DeclareFieldFormat{labelnumberwidth}{[#1] }

\begin{document}

%\newpage
%New things to be included
%
%\begin{itemize}
%\item Dye is dissolved at concentration of 1.5 mM (millimoles per liter) in methanol (see arXiv:1712.08426), which has a refractive index $n=1.33$ for the 532nm laser used here.
%\item At this concentration the dye molecules make basically no difference to the refractive index of the composite liquid (DOI: 10.1063/1.4764565)
%\item People do put chiral molecules in methanol and then do stuff and it's ok (https://doi.org/10.1016/0009-2614(95)00329-3, https://doi.org/10.1073/pnas.93.23.12943)
%\end{itemize}
%%Title of paper
\title{Symmetry-breaking in a condensate of light and its use as a quantum sensor}
\author{Robert Bennett, Yaroslav Gorbachev and Stefan Yoshi Buhmann}

\maketitle

\textbf{{Bose-Einstein condensates (BECs) represent one of the very few manifestations of purely quantum effects on a macroscopic level \cite{Einstein1925,Bose1924}.} The vast majority of BECs achieved in the lab to date consist of bosonic atoms, which macroscopically populate the ground state once a threshold temperature has been reached \cite{Anderson1995}. Recently, a new type of condensate was observed --- the photon BEC \cite{Klaers2010a}, where light in a dye-filled cavity thermalises with dye molecules under the influence of an external driving laser, condensing to the lowest-energy mode. However, the precise relationship between the photon BEC and symmetry-breaking phenomena has not yet been investigated. Here we consider medium-induced symmetry breaking in a photon BEC and show that it can be used as a quantum sensor. The introduction of polarisable objects such as chiral molecules lifts the degeneracy between cavity modes of different polarisations. Even a tiny imbalance is imprinted on the condensate polarisation in a `winner takes it all' effect. When used as a sensor for enantiomeric excess, the predicted sensitivity exceeds that of contemporary methods based on circular dichroism. Our results introduce a new symmetry-breaking mechanism that is independent of the external pump, and demonstrate that the photon BEC can be used for practical purposes.}

The pursuit of Bose--Einstein condensates (BECs) has led to continuing innovation across theory and experiment since its first prediction in 1924 \cite{Einstein1925,Bose1924} and experimental observation in 1995 \cite{Anderson1995}. A remarkable recent development has been the photon BEC, whose first realisation in 2010 \cite{Klaers2010a} brought about a flourishing field of modern research (e.g. \cite{Kirton2013,Kirton2015,VanderWurff2014,DeLeeuw2014a,Nyman2014a,Sela2014}). The condensate of light consists of a macroscopically occupied lowest-energy state of a photon field confined to a cavity, resulting in a central bright spot in the cavity's emission pattern. 
\begin{figure}[t!]\centering
\includegraphics[width=\columnwidth]{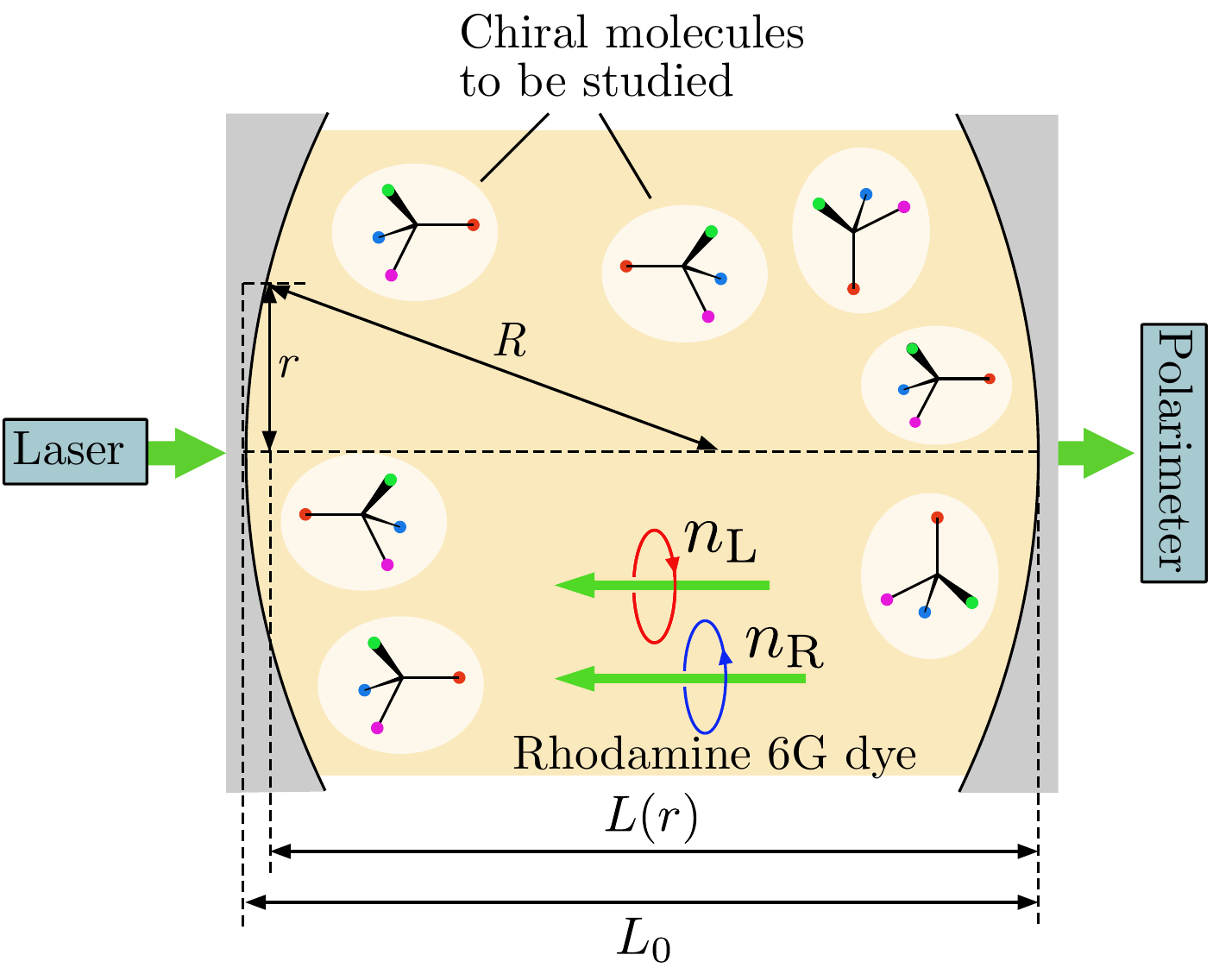}
\caption{Schematic of the physical system considered here. }\label{PhysicalSystem}
\end{figure}

In order to obtain a condensate of light, three conditions must be satisfied. The first of these is that photons should be allowed to exchange energy with each other in a photon-number conserving way. As indicated in Fig.~\ref{PhysicalSystem}, this is achieved by filling a cavity with a dye whose emission and absorption spectra fulfil the Kennard-Stepanov relation \cite{Kennard1918,Stepanov1957}, which allows an effective thermal interaction between the photons via repeated absorption and re-emission. The second is that an effective photon mass should be provided in order to sustain the analogy with atomic condensates, this is given by the curvature of the cavity mirrors which causes the photon dispersion relation to be equivalent to that of massive bosons. Finally, a confining potential is needed---this is naturally provided by the cavity. The resulting system is similar to a laser, one crucial difference being that as the pump power is increased there is a sharp jump in the occupation number of the \emph{lowest} mode, occurring at far below the power required for population inversion.

An aspect of the photon BEC which has not received particular attention, and which uniquely distinguishes it from condensates of massive particles, is its polarisation state. Previous experimental investigations have determined that the photon BEC has a much more well-defined polarisation than a thermal cloud \cite{Greveling2017}. This was investigated in \cite{Moodie2017}, where it was shown that the degree of polarisation differs significantly either side of threshold.

In the mentioned experimental and theoretical investigations, standing-waves modes of differing polarisations have typically been degenerate in energy, meaning that the polarisation degree of freedom is governed entirely by that of driving laser that seeds the condensate. In this work we consider a system, where the polarisation symmetry is broken. We achieve this here by introducing a medium that supports differently-polarised waves with two distinct refractive indices. This lifts the degeneracy between modes of different polarisation, so that a non-trivial polarisation state emerges as an intrinsic feature of the condensate itself. The difference between the two refractive indices provides a simple degree of freedom to tune, with the resulting emission pattern containing signatures of the nature of the medium inside the cavity. 

The symmetry-breaking medium could consist of chiral or anisotropic molecules, or any other system with a polarisation-dependent electromagnetic response. The symmetry-broken condensate will then reveal in-situ information about this system, such as the handedness, orientation or conformational state of molecules. We will use chiral molecules as an example.

We will consider the set-up depicted in Fig.~\ref{PhysicalSystem} where dye molecules fill the space between two gently curved, highly reflecting mirrors and are driven by an external laser. The intra-cavity medium is assumed to be chiral, hence breaking the polarisation degeneracy. After determining the energies and loss rates
of the polarised photon modes, we extend the model by Keeling and Kirton \cite{Kirton2013,Kirton2015}, in which a Jaynes--Cummings-type cavity quantum electrodynamics model was generalised to include the rovibrational states of the dye molecules which are needed to ensure thermalisation, to the case of non-degenerate polarisation. We give a master
equation and rate equations for the coupled light--dye dynamics and use the latter to study the characteristics of the emerging condensate and its dependence on the symmetry-breaking properties of the medium and on the driving laser.

\paragraph{Polarised modes.}

Waves confined in a cavity filled with an optically-active medium of chiral susceptibility $\chi$ are subject to different refractive indices 
\mbox{${n}_\sigma = n \pm \chi$} (\mbox{$\sigma=\mathrm{L},\mathrm{R}$}, see, e.g. \cite{Schaferling2016}) where $n$ is the refractive index of the intra-cavity medium which is assumed to be non-magnetic ($\mu=1$). A cavity of chiral mirrors which reflect left-handed waves into left-handed ones and similarly for right-handed waves, supports standing waves of well-defined circular polarisation. 
 The resulting mode energies
\begin{equation}
 E_{jl\sigma}=\hbar\omega_{jl\sigma}
=\hbar\omega_{j\sigma}^0 +l\hbar\omega_\sigma^\parallel 
\label{eq2}
 \end{equation}
\mbox{($l=0,1,\ldots$)} are \mbox{$(l+1)$}-fold degenerate for spherical mirrors where the off-set 
\mbox{$\hbar\omega_{j\sigma}^0=m_{j\sigma}c_\sigma^2+\hbar\omega_\sigma^\parallel $} is comprised of the normal mode energy and the lateral 
zero-point contribution. In the following, we are going to use a multi-index $\nu=(j,l,\sigma)$ to denote the modes where appropriate. For two highly-reflecting mirrors 
(\mbox{$|r_\sigma|=1-\delta$} with \mbox{$\delta\ll 1$}) and in the paraxial approximation,  the cavity resonances can be shown to be of Lorentzian shape with widths 
\mbox{$\cavityDecay_\nu =2\delta c_\sigma/L_0$} (see Methods section). 

Generalising the works of Keeling and Kirton \cite{Kirton2013,Kirton2015}, the system Hamiltonian can be written in a Jaynes--Cummings \cite{Jaynes1963} form as
\begin{align}
&\hat{H}=\sum_\nu\hbar\omega_\nu \hat{a}^{\dagger}_\nu\hat{a}_\nu
 +\sum_{\nu\alpha}\hbar g_\nu(\hat{a}_\nu\hat{\sigma}^\dagger_\alpha+\hat{a}^{\dagger}_\nu\hat{\sigma}_\alpha)\notag \\
&\!\!+\sum_\alpha\bigl\{\tfrac{1}{2}\hbar\omega_\mathrm{D}\hat{\sigma}^{z}_\alpha
 +\hbar\omega_\mathrm{V}\bigl[\hat{b}^\dagger_\alpha\hat{b}_\alpha+\sqrt{S}\hat{\sigma}^{z}_\alpha\bigl(\hat{b}_\alpha
 +\hat{b}^{\dagger}_\alpha\bigr)\bigr]\bigr\}.
\label{eq4}
\end{align}
Here, the first term is the Hamiltonian of the intracavity field with mode creation and annilhilation operators $\hat{a}_\nu^\dagger$, $\hat{a}_\nu$; 
the second term describes the interaction in the rotating wave approximation  of the cavity modes with the dye molecules indexed by $\alpha$ 
with coupling constants $g_\nu=g_{jl}$ for achiral dye molecules; 
and the last term is the Hamiltonian of the dye molecules with a single electronic transition of frequency $\omega_\mathrm{D}$ described by 
Pauli operators $\hat{\sigma}_\alpha$, $\hat{\sigma}_\alpha^\dagger$, $\hat{\sigma}_\alpha^z$ 
which is coupled to rovibrational states of splitting $\omega_\mathrm{V}$ 
and associated creation and annihilation operators $\hat{b}_\alpha^\dagger$ and $\hat{b}_\alpha$ via a Huang--Rhys factor~$S$.

Applying a polaron transformation, the rovibrational degrees of freedom can be integrated out alongside those modes which are only weakly coupled to the molecules. This leads to a master equation whose Lindblad terms come from pumping of the dye molecules by an external laser ($\pumpRate$), spontaneous decay into field modes that have been integrated out ($\lossRate$) and stimulated emission ($\stimEmission$) as well as absorption ($\stimAbsorp $) from field modes included in the dynamics. From these the coupled rate equations for the mean photon numbers 
\mbox{$\photonN_\nu=\langle\hat{a}_\nu^\dagger\hat{a}_\nu\rangle$} and the excited-state population of the molecules 
\mbox{$p_\mathrm{e}=\langle\hat{\sigma}_\alpha\hat{\sigma}_\alpha^\dagger\rangle$} in a semiclassical approximation can be deduced as
\begin{align}
\dot{\photonN}_\nu &= \cavityDecay \photonN_\nu-\stimAbsorp \photonN_\nu \NMol(1\!-\!p_\mathrm{e}) +\stimEmission(\photonN_\nu\!+\!1)\NMol p_\mathrm{e},\notag \\
\dot{p}_\mathrm{e}&=-\totalLoss p_{e}+\totalGain(1-p_\mathrm{e}) 
\label{eq9} 
\end{align}
with total rates
\begin{align}
\totalLoss&=\!\lossRate+\sum_\nu(l+1)(\photonN_\nu+1)\stimEmission \notag \\ \totalGain&=\!\pumpRate+\sum_\nu(l+1)\photonN_\nu\stimAbsorp
\label{eq11}
\end{align}
where $\NMol$ is the total number of dye molecules.

The main quantities of interest here are photon numbers in the stationary state of the driven-dissipative system. Eliminating the molecular degree of freedom in an adiabatic approximation \cite{Hesten2018}, these can be found as the stationary limit of 
\begin{equation}
\dot{\photonN}_\nu=-\cavityDecay \photonN_\nu
+\NMol\,\frac{\stimEmission(\photonN_\nu+1)\totalGain-\stimAbsorp \photonN_\nu \totalLoss}
{\totalGain+\totalLoss}\,.
\label{eq12}
\end{equation}
Before solving these equations numerically, it is worth discussing the analytical solution which exists in the single-mode case.
%
%\begin{align}
%\photonN_\nu=& \frac{1}{2\cavityDecay( \stimAbsorp +\stimEmission)}\notag\\
%&\times\Bigl(\NMol(\pumpRate\stimEmission-\lossRate\stimAbsorp )-\cavityDecay(\pumpRate+\lossRate+\stimEmission)\notag \\
%&\quad\pm\bigl\{\bigl[\NMol(\pumpRate\stimEmission-\lossRate\stimAbsorp )
%-\cavityDecay(\pumpRate+\lossRate+\stimEmission)\bigr]^2\notag\\
%&\qquad+4\NMol\pumpRate\cavityDecay\stimEmission(\stimAbsorp +\stimEmission)\bigr\}^{1/2}\Bigr)
%\label{eq13}
%\end{align}
%%
For a high-quality cavity with $\cavityDecay\ll\pumpRate,\lossRate,\stimAbsorp ,\stimEmission$, it simplifies to
\begin{equation}
\photonN_\nu=\begin{cases}\displaystyle
0 &\mbox{ if }\tau_{\nu},\\[1ex]
\displaystyle\frac{\NMol(\pumpRate\stimEmission-\lossRate\stimAbsorp )}
{\cavityDecay( \stimAbsorp +\stimEmission)}&\mbox{ if }
\pumpRate\ge\tau_{\nu}.
\end{cases}
\label{eq14}
\end{equation}
which demonstrates the sharp jump in occupation number at a threshold pump frequency given by $\lossRate\stimAbsorp /\stimEmission \equiv \tau_{\nu}$. 
\paragraph{Numerical simulations}

\begin{figure}[h!]\centering
\includegraphics[width=\columnwidth]{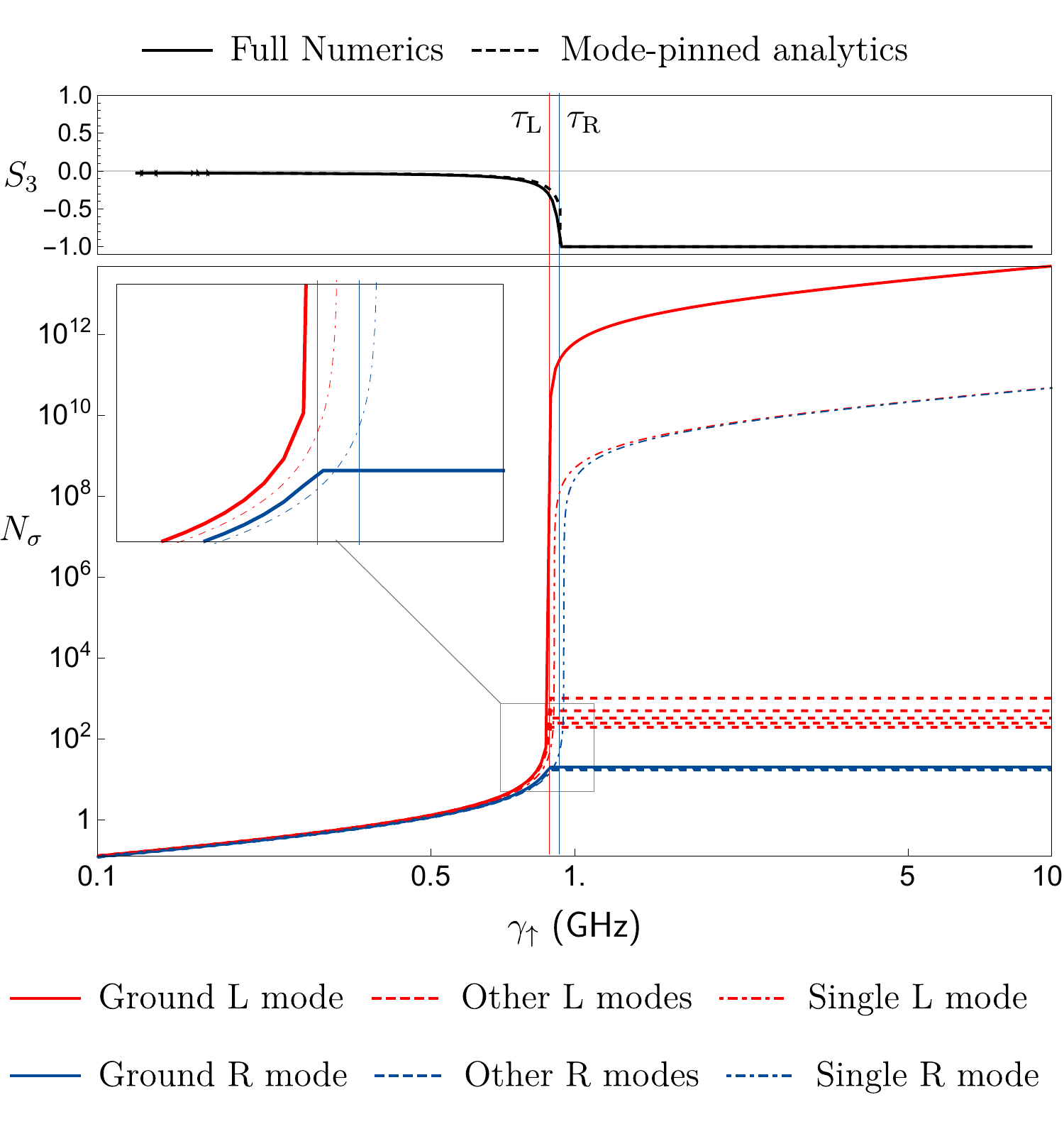}
\caption{Polarisation symmetry breaking in a photonic Bose-Einstein condensate. Lower: Left and right mode occupation numbers $\photonN_\text{L}$ and $\photonN_\text{R}$ as a function of pumping rate $\pumpRate$, with refractive indices  $n_\text{L}=1.3435$, $n_\text{R}=1.3395$. The inset shows a detail of the threshold region where the `pinning' effect discussed in the text can be observed. Upper: Stokes parameter $S_3$ as a function of pumping rate $\pumpRate$, using a full numerical approach (solid line) or a single-model approximation with mode-pinning.}\label{Plot1D}
\end{figure}

The steady-state rate equations \eqref{eq12} are solved with a semi-dynamical approach used in \cite{Hesten2018}. The rates of stimulated emission $\stimEmission$ and absorption $\stimAbsorp$ are derived from the spectrum of rhodamine 6G, which is fitted to experimental data \cite{Nyman2017}.  An example result of the numerical simulations are shown in Fig.~\ref{Plot1D},
where a sharp jump in occupation number for the left-polarised mode is observed as the threshold frequency is crossed. Crucially, this jump is \emph{not} found in the right-polarised mode, meaning that the output signal of the cavity will have a clear `left' polarization. 
 This is a clearly measurable signature of the vectorial nature of the condensate, which cannot be achieved with BECs of massive particles.
The effect can be quantified by investigating the third Stokes parameter:
\begin{equation}
S_3 = \frac{\photonN_\text{R}-\photonN_\text{L}}{\photonN_\text{R}+\photonN_\text{L}}
\end{equation}
 which is shown in the inset of Fig.~\ref{Plot1D}. There the polarization rapidly jumps from ${0>S_3\gtrsim-0.1}$ (slightly left-circularly polarised), to $S_3 = -1$ (strongly left-circularly polarised) as the threshold frequency is passed.

 Fig.~\ref{Plot1D} shows that the single, non-interacting mode approximation does not capture the essential physics of the symmetry-breaking effect. This is to be expected, as the single-mode approach neglects transfer of energy between left- and right-circularly polarised modes, which is the mechanism that allows our system to pick out one polarisation. Nevertheless, its qualitative behaviour can be described by a `pinning' process, where whichever polarisation has a lower condensation threshold as pump power is increased becomes macroscopically occupied, with the other one staying pinned to whatever occupation it had when the \emph{other} polarisation hit its threshold (even if the pump power is far above its own threshold). This process is the origin of the `winner takes it all' character of the system, and as shown in the upper panel of Fig.~\ref{Plot1D} this simple approach gives good approximations for the polarisation of the output beam.

\begin{figure}[h!]\centering
 \includegraphics[width=\columnwidth]{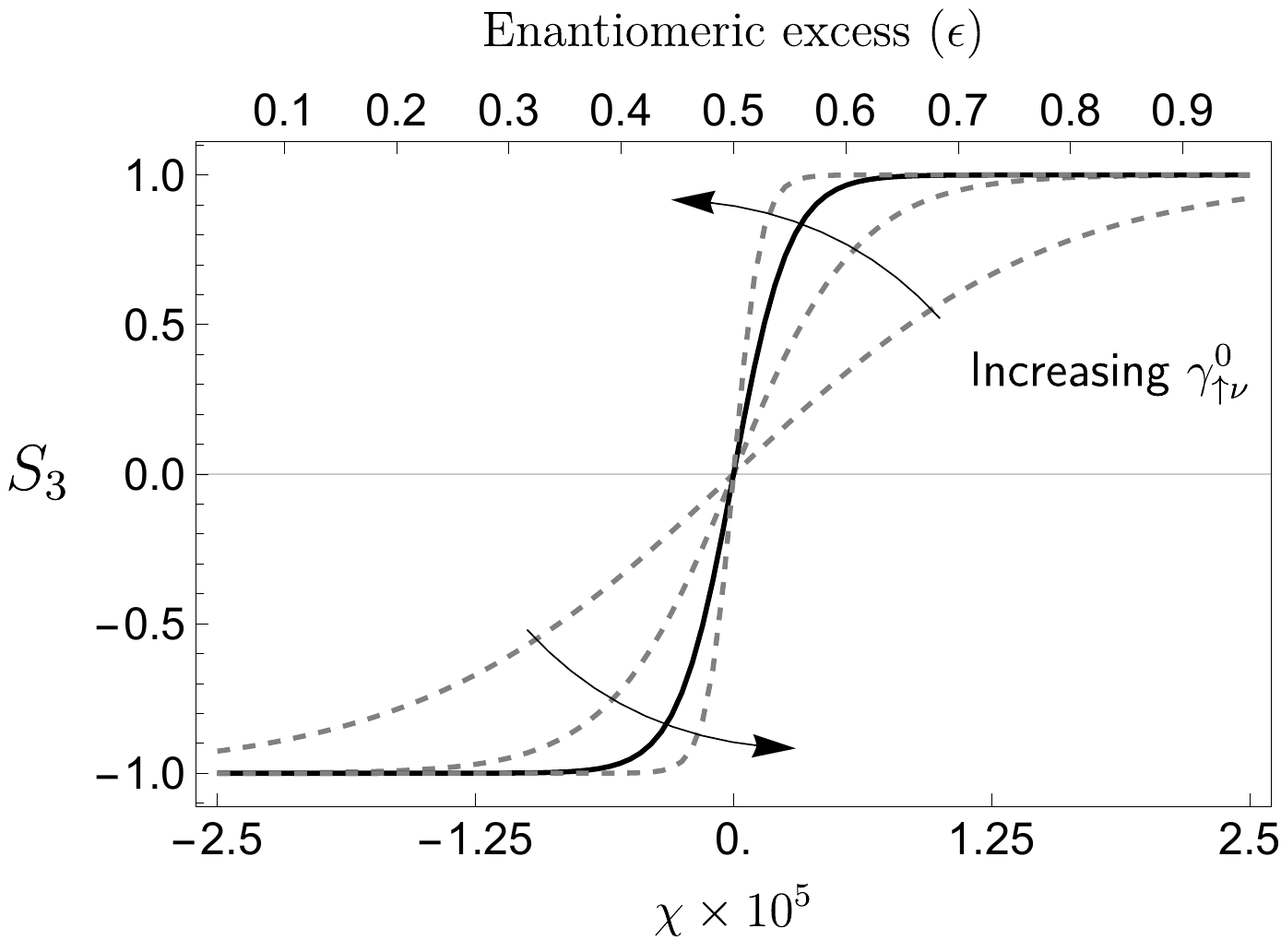}
    \caption{Use of the photonic Bose-Einstein condensate as a detector for enantiomeric excess. Condensate polarisation $S_3$ as a function of refractive index difference and chiral parameter $\chi$ (multiplied by $10^5$) and enantiomeric excess $\epsilon$. The solid line corresponds to the parameters shown in Tab. \ref{ParamsTable}, while the dotted lines have had $\stimAbsorp^0$ scaled by factors 0.5, 2 and 10 in the order indicated. }\label{VaryEpsPlot}
    \end{figure}

\begin{figure}[h!]\centering
 \includegraphics[width=\columnwidth]{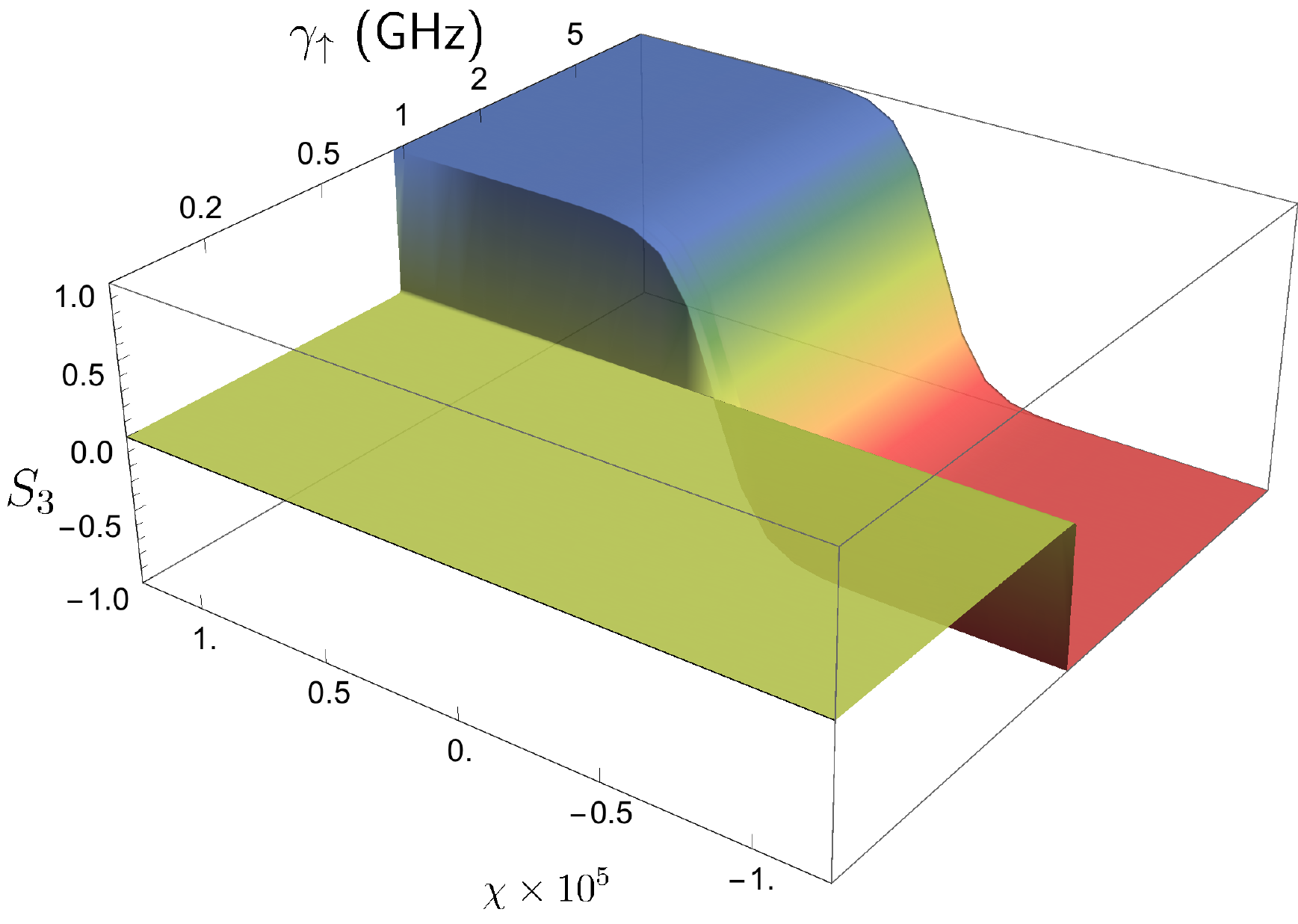}
    \caption{Stokes parameter $S_3$ as a function of chiral parameter $\chi$ and pumping strength $\pumpRate$, using the parameters shown in Tab. \ref{ParamsTable}.}\label{Results3DS3}
\end{figure}

\paragraph{Sensitivity} Let us assess the ability of the proposed system to detect realistic chiral imbalances. As shown in the Methods section, the magnitude of the chiral parameter $\chi$ can be calculated through
\begin{equation}\label{ChiValue}
|\chi| = 2.14 \times 10^{-8} \theta m_u \epsilon \alpha^2
\end{equation}
where $m_u$ is the mass of each chiral molecule in atomic mass units $u=1.66 \times 10^{-27}$kg, $\theta$ is the tabulated specific rotation of the chiral molecules with volume fraction $\alpha$ and $\epsilon$ is the enantiomeric excess, defined as the absolute value of the difference between the volume fractions of the two enantiomers. Precise determination of $\epsilon$ is the subject of considerable pharmacological interest \cite{Caldwell2007a}. Considering a solute of glucose (mass 180$u$, $\theta=44\degree$) with volume fraction $\alpha = 40\%$,  one finds $\chi = 2.7 \times 10^{-5}\epsilon $.  The enantiomeric excess is shown as a percentage on the upper axis of Fig \ref{VaryEpsPlot}, 
where $\chi$ is varied for a single pump frequency well above threshold ($\pumpRate=10$GHz). The gradient ${dS_3}/{d\epsilon}$ at $\epsilon = 0.5$  is approximately 16, showing that a state-of-the-art cryogenic enantiosensitivity of 0.02 \cite{Patterson2014} can be detected as a shift of around 0.32 in the Stokes parameter $S_3$. The accuracy of the Stokes parameter measurements already carried out in the first (and so far only) photon BEC polarisation experiments \cite{Greveling2017} was around $\pm 0.2$. Figure \ref{VaryEpsPlot} also demonstrates how the sensitivity of the device and its operating window (i.e. the range and slope of the circular polarisation as a function of enantiomeric excess between the two plateau regions) can be modified by varying the dye-molecule response $\stimAbsorp$ at the frequency of the laser-cavity system. Finally, Fig.~\ref{Results3DS3} 
shows a summary of the ideas presented here, where both the refractive index difference and the pumping strength are varied simultaneously. While the system is unpolarised below threshold, a unique circular polarisation emerges above threshold that is governed by the enantiomeric excess. We note that any type of chiral molecule or structure which can be dissolved a solvent alongside rhodamine 6G could be investigated in this way. Taking methanol as an example solvent as above, these include for instance $\alpha$-methyl-2-naphthalenemethanol \cite{Al-Rabaa1995} and phospholipid tubules \cite{Spector1996}. \\
 
 \paragraph{Summary and conclusions.} The photon BEC is a remarkable new physical system whose applications are only beginning to be explored. Here we have proposed a polarisation-symmetry breaking mechanism that can serve as a sensitive probe for magnetoelectric properties of liquid media. We have illustrated this for the case of chiral molecules, where the condensate can be used as an extremely sensitive probe of the enantiomeric excess of a liquid medium at room temperature. These operating conditions mean that the pharmacological applications of the completely new type of chiral sensor proposed here could be widespread. Our analysis applies to any photon BEC with two different types of mode, such as for instance birefringent media where the horizontal-vertical polarisation can serve as a diagnostic tool for detecting the orientation of anisotropic molecules or, by proxy, very weak electric or magnetic fields inducing this orientation. Signatures of conformational chemical reactions could then be observed in real-time. A possible future direction to be explored is the interplay of the mode geometry with the spatial distribution of symmetry-breaking probe molecules in analogy with cavity-QED setups for position monitoring and control \cite{Pinkse2000}. In a broader context, this shows that the photon BEC is ready for an explosion of practical applications in the same way the laser was after its first realisation in the lab. \\
 
{\noindent \sffamily \textbf{Acknowledgements}} The authors thank Henry Hesten, Robert Nyman, Andreas Buchleitner and Akbar Salam for fruitful exchanges. R.B. acknowledges financial support by the Alexander von Humboldt Foundation, Y.G. and S.Y.B. thank the Deutsche Forschungsgemeinschaft (grant BU 1803/3-1476). R.B. and S.Y.B. both acknowledge support from the Freiburg Institute for Advanced Studies (FRIAS).\\
\newpage
{\noindent \sffamily \textbf{Competing financial interests:}} Patent application, \\
Applicant: University of Freiburg\\
Inventors: S. Y. Buhmann, R. Bennett, and Y. Gorbachev\\
Submitted to: United States Patent and Trademark Office, application no. 62/818,505, 14/03/2019.  \\
Aspect of the work covered: Use of the photon BEC as a sensitive detector of enantiomeric excess \\ 
{\noindent \sffamily \textbf{Author contributions:}} S.Y.B. conceived the project, R.B. implemented the numerical simulations. S.Y.B and R.B. wrote the manuscript, while all authors contributed to analytical calculations, discussion and interpretation of the results.   \\
{\noindent \sffamily \textbf{Author to whom correspondence and material requests should be addressed}}: Stefan Buhmann (stefan.buhmann@physik.uni-freiburg.de) 
 
\printbibliography 

%\subsection{Operating Procedure}
%
%
%\subsection{Modes of operation, applications}
%
%The device is intended to have at least two modes of operation. The first of these is where the presence of any chiral molecules in an otherwise achiral substance would be detected. The second mode would be where small deviations from an ideal racemic mixture would be detected in real time by continuously measuring the polarisation of the output light. This could be fed into a feedback mechanism where racemicity would be restored by addition of the appropriate enantiomer through the input channel. 
%
% \section{Novelty}
%
%A search of the literature concludes that no method for determination of enantiomeric excess using a BEC (of any type) has ever been proposed or implemented.
%
%\section{Advantages}
%
%The solution presented here has the following advantages over the state-of-the-art:
%
%\begin{enumerate}
%\item The device works at room temperature, avoiding the complicated and expensive process of cooling to cryogenic temperatures.
%\item The device does not require the chiral molecules to be in the gas phase, rather having them immersed in a solvent
%\item The device is extremely sensitive to even the smallest changes in directional refractive index due to the sharp nature of the BEC phase transition, meaning that it can outperform the best cryogenic methods of enantiomeric excess determination. 
%\item The device can detect changes in the enantiomeric excess of a solution in real time. 
%\end{enumerate}
%

%------------------------------------------------

\newpage 
\section*{Methods}

\paragraph{Determination of mode energies.}

To determine the energies of the circularly polarised standing-wave modes in the cavity shown in Fig.~\ref{PhysicalSystem}, we generalise the procedure given in Ref.~\cite{Klaers2010a}. The cavity is assumed to be formed by two curved chiral mirrors of identical radius $R$. Denoting the mirror separation along the optical axis by $L_0$, the cavity length at a lateral distance $r$ away from the optical axis is \mbox{$L(r)=L_0-2\bigl(R-\sqrt{R^2-r^2}\bigr)$}. Separating the wave vector 
\mbox{$k=\sqrt{k_z^2+k_\parallel^2}$} into its normal and lateral components $k_z$ and $k_\parallel$, respectively,
using the dispersion relation \mbox{$E=\hbar kc_\sigma$} (\mbox{$c_\sigma=c/n_\sigma$}: speed of light within the chiral medium with $c$ denoting the vacuum speed of light) and noting that the normal component of a standing-wave mode has to fulfil the resonance condition \mbox{$k_z=\pi j/L(r)$} (\mbox{$j=1,2,\ldots$}), the associated energy reads
\begin{align}
 E_{j\sigma}(r,k_\parallel)
&=\hbar c_\sigma\,\sqrt{\frac{j^2\pi^2}{L^2(r)}+k_\parallel^2}\notag \\
&\approx   \frac{\pi \hbar c_\sigma j}{L_0}+\frac{\hbar c_\sigma L_0k_\parallel^2}{2\pi j}+\frac{\pi\hbar c_\sigma jr^2}{2L_0^2R}
\label{eq1}
 \end{align}
in the paraxial approximation (\mbox{$r\ll R$}, \mbox{$k_\parallel r\ll 1$}). The lateral mode structure is hence that of a two-dimensional harmonic oscillator with an effective photon mass \mbox{$m_{j\sigma}=\pi\hbar j/(c_\sigma L_0)$} and a lateral frequency \mbox{$\omega_\sigma^\parallel=c_\sigma/\sqrt{L_0R}$}.

\paragraph{Cavity decay}

To describe the dynamics of these modes for imperfectly reflecting mirrors, we also need to determine their cavity leakage or decay constants $\cavityDecay $. To this end, we consider the electromagnetic Green's tensor $\ten{G}(\vect{r},\vect{r},\omega)$ which is directly related to the spectral energy density 
\mbox{$\rho=(\hbar/\pi)(\omega^2/c^2)\trace\operatorname{Im}\ten{G}$} of the electromagnetic field inside the cavity. When neglecting the mirror curvature, the cavity Green's tensor is given by
\begin{align}
\ten{G}^{(1)}&(\vect{r},\vect{r},\omega) =\frac{\mi}{8\pi^2}\int\frac{\dif^2k_\parallel}{k_z}\sum_\sigma \frac{1}{D_\sigma}\notag \\
&\times \Bigl[r_\sigma^2\me^{2\mi k_zL_0}(\vect{e}_{\sigma+}\vect{e}_{\sigma+}+\vect{e}_{\sigma-}\vect{e}_{\sigma-})\notag\\
&+r_\sigma\bigl(\me^{2\mi k_z z}\vect{e}_{\sigma+}\vect{e}_{\sigma-}+\me^{-2\mi k_z z}\vect{e}_{\sigma-}\vect{e}_{\sigma+}\bigr)\Bigr]
\label{eq3}
\end{align}
where $r_\sigma$ are the reflection coefficients of the cavity mirrors, $\vect{e}_{\sigma\pm}$ are the polarisation units vectors for the respective $\sigma$-polarised plane waves traveling in the positive or negative $z$-direction, respectively and \mbox{$D_\sigma=1-r_\sigma^2\me^{2\mi k_zL_0}$}. For two well-reflecting mirrors 
(\mbox{$|r_\sigma|=1-\delta$} with \mbox{$\delta\ll 1$}) and in the paraxial approximation, the denominators in the vicinity of the cavity resonances read
\mbox{$D_\sigma\propto(\omega-\omega_{j\sigma}^0)+\mi\cavityDecay /2$} with $\cavityDecay= 2\delta c_\sigma/L_0$, showing that the cavity resonances are of the Lorentzian shape used in the main text. 

\paragraph{Deriving the rate equations}

Beginning from the Hamiltonian \eqref{eq4}, we apply a polaron transformation \mbox{$\hat{H}\mapsto\hat{U}^\dagger\hat{H\hat{U}}$} with 
\mbox{$\hat{U}=\exp\big[\sum_\alpha\sqrt{S}\hat{\sigma}^{z}_\alpha(\hat{b}_\alpha-\hat{b}_\alpha^\dagger)\big]$} \cite{Marthaler2011}, after which the Hamiltonian assumes the form
\begin{align}
\hat{H}=&\sum_\nu\hbar\omega_\nu \hat{a}^{\dagger}_\nu\hat{a}_\nu
 +\sum_\alpha\bigl(\tfrac{1}{2}\hbar\omega_\mathrm{D}\hat{\sigma}^{z}_\alpha
 +\hbar\omega_\mathrm{V}\hat{b}^\dagger_\alpha\hat{b}_\alpha\bigr)\notag \\
&+\sum_{\nu\alpha}\hbar g_\nu(\hat{a}_\nu\hat{\sigma}^\dagger_\alpha\hat{D}_\alpha
+\hat{a}^{\dagger}_\nu\hat{\sigma}_\alpha\hat{D}_\alpha^\dagger)
\label{eq5}
\end{align}
 The first term is the Hamiltonian of the intracavity field with mode creation and annihilation operators $\hat{a}_\nu^\dagger$, $\hat{a}_\nu$.
The second term is the free Hamiltonian for a dye molecule with a single electronic transition (frequency $\omega_\mathrm{D}$, described by 
Pauli operators $\hat{\sigma}_\alpha$, $\hat{\sigma}_\alpha^\dagger$, $\hat{\sigma}_\alpha^z$ ), and a ladder of rovibrational levels (frequency splitting $\omega_\mathrm{V}$, described by creation and annihilation operators $\hat{b}_\alpha^\dagger$ and $\hat{b}_\alpha$). The final term describes the mutual interactions of photons with the electronic and rovibrational transitions through the coupling constants $g_\nu$ and the Huang-Rhys factor $S$. We can then integrate out the rovibrational degrees of freedom together with those field modes which are only weakly coupled to the molecules to arrive at a master equation \cite{Kirton2013,Kirton2015}
\begin{align}
\dot{\hat{\rho}}=&-\frac{\mi}{\hbar}\bigl[{\hat{H}_0,\hat{\rho}}\bigr]\notag \\
&-\biggl(\sum_\nu \frac{\cavityDecay}{2}\mathcal{L}[\hat{a}_\nu]
+\sum_\alpha\biggl\{\frac{\pumpRate}{2}\mathcal{L}\bigl[\hat{\sigma}_\alpha^\dagger\bigr]+\frac{\gamma_{\downarrow}}{2}\mathcal{L}[\hat{\sigma}_\alpha]\biggr\} \notag\\
&+\sum_{\nu\alpha}\bigg\{\frac{\stimAbsorp }{2}\mathcal{L}\bigl[\hat{a}_\nu\hat{\sigma}_\alpha^\dagger\bigr]\biggr\} 
+\frac{\stimEmission}{2}\mathcal{L}\bigl[\hat{a}_\nu^\dagger\hat{\sigma}_\alpha\bigr]\biggr)\hat{\rho} 
\label{eq6} 
\end{align}
where $\hat{H}_0$ describes the Hamiltonian dynamics of the electronic dye transition and the remaining cavity modes (which play no further role here). The notation in the Lindblad terms \mbox{$\mathcal{L}[\hat{O}]\hat{\rho}=\{\hat{O}^\dagger\hat{O},\hat{\rho}\}-2\hat{O}\hat{\rho}\hat{O}^\dagger$} is explained in the main text.  The rates for the stimulated emission and absorption processes read 
\mbox{$\stimEmission=2\operatorname{Re}K_\nu(\omega_\mathrm{D}-\omega_\nu)$} and
\mbox{$\stimAbsorp =2\operatorname{Re}K_\nu(\omega_\nu-\omega_\mathrm{D})$} with
\begin{equation}
K_\nu(\omega)=g_\nu^2\int_0^\infty \dif t\, \bigl\langle\hat{D}_\alpha(t)\hat{D}_\alpha(0)\bigr\rangle 
 \me^{-(\pumpRate+\lossRate) t/2}\me^{-\mi\omega t}.
\label{eq7} 
\end{equation}
Note that the spatial variation of both the coupling constant $g_\nu$ and the rates $\lossRate$, $\stimEmission$
and $\stimAbsorp $ have been neglected, so that these quantities are equal for all dye molecules. The master equation \eqref{eq6} allows semiclassical rate equations for the mean photon numbers and the excited-state population of the molecules to be derived, these are Eqs.~\eqref{eq9}  in the main text.

\begin{figure}[h!]\centering
 \includegraphics[width=\columnwidth]{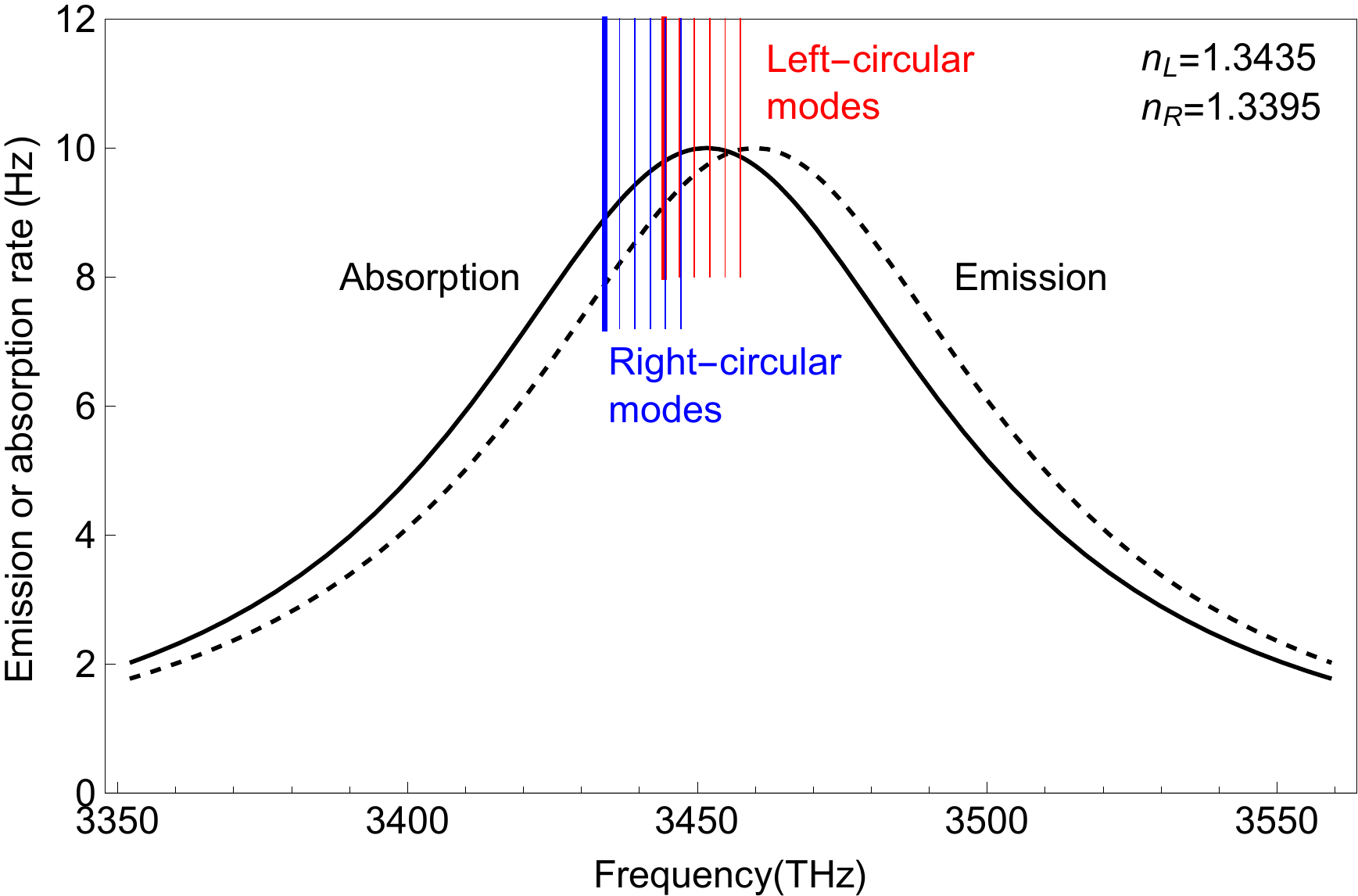}
    \caption{Emission and absorption spectrum. Every tenth mode ($l=0,10,20...$) is shown, while the lowest mode for each polarisation is shown as a thick line.}\label{DyeSpectrum}
\end{figure}

\paragraph{Numerical simulations} We model the response of the dye molecules via the following functions 
\begin{align}
\stimEmission&=\frac{\dyePeakBroadening ^2 \stimEmission^0}{\frac{\dyePeakBroadening ^2}{4}+(\dyeCenter+l \omega_\sigma^\parallel -\peakSplitting)^2}, \notag \\
 \stimAbsorp&=\frac{\dyePeakBroadening^2 \stimAbsorp^0}{\frac{\dyePeakBroadening ^2}{4}+(\dyeCenter+l \omega_\sigma^\parallel +\peakSplitting)^2}
\end{align}
with parameters listed in Table \ref{ParamsTable}. 
  \begin{table}[h]
  \centering
  \begin{tabular}{*4l}
    \toprule
\multicolumn{2}{l}{Dye and solvent parameters}  &   \multicolumn{2}{l}{Cavity parameters}   \\ 
    \midrule
$\dyePeakBroadening $ & 50 THz \cite{Nyman2017}  & $R$ & 1\,m \cite{Klaers2010a} \\ 
   $\lossRate,g_\nu$ & 1 GHz \cite{Kirton2013}  & $L_0$ & 1.46 $\mu$m \cite{Klaers2010a} \\ 
$\stimEmission^0,\stimAbsorp^0$ &10 Hz \cite{Nyman2017}  & $\NMol$ & $10^9$ \cite{Kirton2013} \\ 
        $\peakSplitting$ &4.18 THz  \cite{Nyman2017}  &     $\cavityDecay$ & 100 MHz  \cite{Kirton2013} \\ 
          $\dyeCenter$ &3 456 THz \cite{Nyman2017}&    &$\rotatebox[origin=c]{180}{$\Lsh$}(|r_\sigma|= 0.99$)  \\
                    $n$ & 1.34 \cite{Sanchez-Perez2012} &  $j$    & 7 \cite{Klaers2010a}  \\
    \bottomrule
  \end{tabular}
              \caption{Parameters used in the numerical simulations. The cavity parameters are taken directly from the experiment \cite{Klaers2010a}, the solvent and dye spectral parameters are fitted to measured data \cite{Sanchez-Perez2012,Nyman2017}, and the remainder are taken from theoretical work \cite{Kirton2013} describing the experiment \cite{Klaers2010a}.  }\label{ParamsTable}
  \end{table}
The resulting emission and absorption spectrum is shown in Fig.~\ref{DyeSpectrum}, 
where the $j=7$ manifold is selected, which is assumed throughout this work.

\paragraph{Chiral sensitivity.}

The value of the chiral parameter $\chi$ in terms of the wavelength $\lambda$ of light incident on a sample of molecular number density $\rho$ is \cite{Schaferling2016};
\begin{equation}\label{ChiExp}
|\chi | = \frac{|\Theta| \alpha \rho m \lambda}{2\pi} \;. 
\end{equation}
Here, $\alpha$ is the volume fraction of chiral molecules of mass $m$ in the whole solution, and $\Theta$ is the specific rotation given by
\begin{equation}\label{ThetaExp}
\Theta = \alpha(\alpha_\text{R} - \alpha_\text{L}) \theta \cdot 0.01 \, \text{m}^2/\text{kg} \qquad (|\alpha_\text{R} - \alpha_\text{L} |= \epsilon)  
\end{equation}
where $\theta$ is the tabulated value for a particular molecule (often referred to itself as the specific rotation), and $\alpha_\sigma$ are the volume fractions of each enantiomer within the chiral component of the solution ($\alpha_\text{L}+\alpha_\text{R} =1$), so that the absolute value of their difference is the enantiomeric excess $\epsilon$. The angle $\theta$ parameterises how much a particular chiral molecule rotates an incoming beam of circularly polarised light, independent of the concentration and propagation length. The number density of the methanol which fills the cavity in the proposed device is $\rho=1.488 \times 10^{28}\text{m}^{-3}$, and its operating frequency of 3450 THz corresponds to a wavelength of $546$nm. Substituting these values into Eq.~\eqref{ChiExp} via Eq.~\eqref{ThetaExp} one arrives at Eq.~\eqref{ChiValue}.

\newpage

\end{document}